\documentstyle[prd,aps,floats]{revtex}
\input{psfig.tex}

\newcommand{\beq}{\begin{equation}}
\newcommand{\eeq}{\end{equation}}
\newcommand{\bea}{\begin{eqnarray}}
\newcommand{\eea}{\end{eqnarray}}

\def\Mp{M_{\rm Pl}}
\def\mpcunit{m_{Pl}^2 {\rm Mpc}^{-2}}

\def\ten#1{\times 10^{#1}}
\def\phii{{\phi_i}}
\def\lambdai{\lambda_i}
\def\rhoi{{\rho_i}}
\def\rhocrit{\rho_{crit0}}

\begin{document}

\title{Toward a Possible Solution to the Cosmic Coincidence Problem}
\author{Kim Griest\\
Physics Department 0319\\
University of Calfornia, San Diego\\
La Jolla, CA 92093}
\maketitle

\begin{abstract}
We suggest a paradigm that might allow for a non-anthropic solution
to the cosmic coincidence problem of why the density of vacuum energy
and matter are nearly equal today.
The fact that the half life of
Uranium 238 is near to the age of the solar system is not considered
a coincidence since there are many nuclides with a wide range of half lives
implying that there is likely
to be some nuclide with a half life near to any given time scale.
Likewise it may be that the
vacuum field energy causing the universal acceleration today is just
one of a large ensemble of scalar field energies, which have dominated
the Universe in the past and then faded away.
Predictions of the idea include: the current density of vacuum energy is 
decreasing, the ratio of vacuum pressure to vacuum density, $w$, is
changing and not equal to $-1$, there were likely periods of vacuum
domination and acceleration in the past and may be additional periods
in the future, and the eventual sum of all scalar field vacuum densities
may be zero.
\end{abstract}

\section*{Introduction}
The discovery that the Universe is accelerating 
\cite{perlmutter:98,reiss:98}, probably due to
dominance of vacuum energy, has pushed two uncomfortable fine-tuning
problems to the forefront of the physics community.
The first problem\cite{weinberg:89} 
is the that scale of this vacuum energy density, 
(.002 eV)$^4$
is vastly different from the MeV to Planck scale particle physics 
vacuum energies that must sum to give this number.
In the past, one hoped that some unknown symmetry principle would require
these energies to ultimately sum to zero, but the existence of a non-zero
sum naively dashes these hopes.
The second fine-tuning problem occurs because traditional vacuum energy density
does not change with time while matter and radiation density change rapidly
as the Universe expands.
At early times the Universe was radiation and then matter dominated, and
only recently, at $z\sim 1.7$, became vacuum dominated.
The importance of this ``cosmic
coincidence" is that in the normal scenario, once vacuum energy becomes
dominate, it stays dominate, and if this had occurred at any earlier
epoch, the evolution of the Universe would have been completely different
and most likely we wouldn't be here to discuss it.  
Currently, there is no convincing fundamental physics idea for why
vacuum dominance happened only recently, and this has led many workers
\cite{vilenkin:01,garriga:01,stewart:00}
to conclude that some sort of anthropic principle must be at work.
The anthropic idea\cite{weinberg:89} is that there is an ensemble of 
universes with different values of
the vacuum energy, most of which do not allow life to develop.  Therefore
the cosmic coincidence is ``explained" by saying the existence of
intelligent life selects only those values of vacuum energy density near that
which has been measured in the SN Ia observations.
Many non-anthropic ideas have been proposed, which reduce the coincidence,
usually by having the vacuum energy ``track" the matter density in some
way so that the ratio is not so large\cite{caldwell:98},
but all of these have been criticized as in fact involving fine-tuning
in some way \cite{vilenkin:01,garriga:01,stewart:00}.  
At this point, some workers have concluded that only
anthropic ideas have a chance of explaining the cosmic 
coincidence \cite{garriga:01}.

The goal of this paper is to propose a new class of solutions to the cosmic
coincidence problem which is not anthropic and which may allow removal
of fine-tuning.
The idea presented here is similar to, but somewhat more general than,
that of the tracking oscillating potential model of 
Dodelson, Kaplinghat, and Stewart \cite{dodelson:01}.
This class of models also makes it possible to change the current
cosmological constant problem from why the vacuum energies 
sum to a small number, back to the older problem of why the vacuum
energies sum to zero.  We do not have anything to say about the
basic cosmological constant problem of why this sum should be zero in
first place.  
To understand our class of potential solutions an analogy may be helpful.

\section*{An Analogy}

Suppose we demand a non-anthropic solution to the cosmic coincidence problem.
What would such a solution look like?  Physics offers several examples
of potential cosmic coincidences that we do not consider as such.
For example, U238 is a common radioactive substance with a half life of
4.5 Gyr, almost exactly the age of the solar system.  
The reason that this is not a cosmic coincidence is clear in this case.
There are thousands of nuclides with half lives ranging over a enormous
range of time scales, from microseconds (e.g. U222, $\tau=1 \mu$s), 
to seconds (e.g. U226, $\tau = 0.5$s), to days (e.g. U231, $\tau=4.2 $d),
to millennia (e.g. U233, $\tau=1.6\ten{5}$ yr) to the age of the Universe
(e.g. U238), to $10^{20}$ years (e.g. Se82, $\tau=1.4\ten{20}$ yr),
and up.

Thus there is no surprise that for any given time scale, such
as the current age of the Solar System, there is some nuclide which is decaying 
on just this time scale.  
Note that if U238 were the only nuclide in existence and everything were
made of it, then there would
be a cosmic coincidence problem quite similar to the vacuum energy cosmic
coincidence problem.  Also, if the range of radioactive decay constants
were not exponentially distributed over such a wide range of time scales
then again even with hundreds of nuclides, finding one with $\tau=4.6$ Gyr
would be unlikely.
In this example, it is the exponential sensitivity of radioactive
decay to the nuclear wavefunction that allows a small change, such as the 
addition of one neutron, to make a large change in the half life.

Thus we are inspired to seek a possible class of models to 
solve the vacuum energy cosmic coincidence
by positing not just one scalar field
whose vacuum energy is making its appearance today, but an ensemble of
fields, whose vacuum energies span an exponentially large range of 
energy densities,
some of which have dominated the Universe for short periods of times at
many time scales in the past.  We note that we have no a priori reason
for the existence of such an ensemble of vacuum fields, but are using
the cosmic coincidence problem as a clue that such an ensemble might
exist.

In this class of models, the specific field (or fields) responsible for 
the current acceleration 
is not special, but just happens to be the one dominant at this time.
We note that in this context, the purported inflaton field responsible
for cosmic inflation in the very early Universe could be just another
one (or more than one) of these fields.

This type of solution to the cosmic coincidence problem
makes some interesting predictions:

\begin{itemize}

\item
The vacuum energy is not a cosmological ``constant".  The current
	phase of acceleration is temporary and will eventually finish;
	$w = p/\rho$ is not $w=-1$ exactly.

\item
There probably were several periods of acceleration and vacuum dominance
	in the past, followed by radiation and/or matter domination, and
	finally vacuum dominance again today.

\item
There will likely be additional periods of acceleration in the future.
	Thus predicting the ultimate fate of the Universe will not
	be possible without an understanding of the origin of all these fields
	and their vacuum energies.

\item
The sum of all these changing vacuum energies may well eventually be 
zero; that is, the minimum of the potential of all 
these fields may be a zero
	that we are evolving toward.
	Just as even very long lived nuclides will eventually decay, it
	is thus possible that the actual cosmological ``constant" is zero,
	and we are just part way there.  We offer no suggestion 
	here as to why the sum should be exactly zero, just note that in this
	scheme it is possible.  
	However, just as some nuclides are stable,
	it is also possible that the final cosmological constant is not zero.
\end{itemize}
We note that the tracking oscillating model of Dodelson, Kaplinghat,
and Stewart \cite{dodelson:01} is similar to our class of models.  
It uses a single scalar field with a potential with many 
wiggles and also predicts
many periods of acceleration in the past.  It addresses the cosmic
coincidence problem in a way similar to ours.  One difference
is that the single potential must be 
quite complicated to allow such behavior, 
and it seems interesting and somewhat more generic and
flexible to consider an ensemble of scalar fields with simple potentials.

\section*{Constraints}
There are several observations that constrain vacuum energy density
in the Universe and which any model of an accelerating
Universe must satisfy.  Our model contains several fields and so
differs somewhat from single field quintessence/dark energy 
models \cite{caldwell:98}.  However,
given that we do not have a specific
model, we will not attempt an accurate tallying of the constraints,
but only mention a few constraints that should be considered.

First, if the density of scalar field energy, $\rho_{vac}$, is more than a few
percent of the radiation density, $\rho_{rad}$, at 
the time of big bang nucleosynthesis (BBN),
the universal expansion rate will increase enough to make a significant
difference in the predicted abundances of helium and deuterium.
Since these abundances are fairly well measured, we demand
$\rho_{vac}/\rho_{rad} < .02$ when $10^8 < z < 10^{10}$\cite{yahiro:01}.

Second, the equation of state $w$ must be sufficiently negative for
the scalar field energy solutions to match the current 
type Ia supernova measurements
and be consistent with the cosmic microwave background (CMB) and 
large scale structure measurements.  We require
$w< -0.79$ (95\% CL) \cite{bean:02}. 
In addition we can directly calculate the absolute
magnitude-redshift relation and compare to the supernovae reported in
Perlmutter, et al. \cite{perlmutter:98}, demanding that, within errors, it be 
as good a fit as a cosmological constant model.
Also, since the growth of large scale structure can be reduced
by periods of vacuum dominance, or near vacuum dominance, there are several
observational constraints on the power spectrum that are possible.
For simplicity we will merely
calculate the linear growth factor from the time of matter-radiation
equality until the present,
and compare it to that expected from a cosmological constant model.

Next, additional vacuum energy at the time of photon decoupling can
shift the well measured acoustic peaks in the CMB, providing the constraint
$\rho_{vac}/\rho_{other} < 0.64$, at $z\approx 1100$
\cite{bean:01,hu:95}
where $\rho_{other}$ includes dark matter, radiation, and baryons.  
To preserve the peak positions, we can also demand that
the angular diameter distance to surface of last scattering at $z \approx 1100$
not be too different from that implied by a cosmological constant model.

We easily find models in which these constraints are satisfied, basically
by chosing the field content and parameters such that
none of the additional scalar fields is very important during 
either the decoupling or nucleosynthesis epochs, and such that
the field that dominates today has $w$ in an appropriate range.  
From one point of view it thus requires 
fine tuning of many parameters to satisfy these constraints
and so these restrictions are a weakness of our idea.  
However, from another point of view,
these constraints and the data from which they derive
are actually just measurements of the initial scalar field content
and parameters.  That is, had these values been different, then the
present the values of our current cosmological parameters and the
the positions of the CMB peaks would now be different.  

\section*{Example model}
Since this class of solutions is motivated purely by solving the
cosmic coincidence problem, there are no restrictions on the types
of fields or forms of potentials that may be used.  As a simple example
consider an ensemble of $N$ scalar fields $\phii$, $i = 1,...,N$ 
that do not interact with ordinary matter or each other, 
and which have potentials of the form
\beq
	V_i(\phii) = \lambda_i \phii^{\alpha_i}.
\label{eqn:potentialeqn}
\eeq
The total scalar field potential is then just $V = \sum_i V_i(\phii)$.
To first approximation (ignoring gravity, finite temperature
effects, parametric resonance, etc.)
each field is governed by the standard equations:
$\ddot{\phii} + 3 H \dot{\phii} + V_i^\prime(\phii) = 0$,
$H^2 = \frac{1}{3\Mp^2}[\rho_{\rm other}+\frac{1}{2}\sum_i\dot{\phii}^2+
\sum_i V_i(\phii)]$, where
$H(z) \equiv \dot{a}/a$, and 
the Planck mass is
$\Mp = {(8 \pi G_N)}^{-1/2} = 2.44 \times
10^{18} {\rm GeV}$, and where $a$ is the scale
factor of the Universe, $\rho_{\rm other}$ is
the total energy density of the other contributing fields, 
the dot represents differentiation with respect to time,
and the prime indicates differentiation with respect to $\phi_i$.
Note that with $\alpha=4$, the
$\lambdai \phii^{\alpha_i}$ form for the potential is that
used in chaotic inflation\cite{linde:55}, and if $\phii$ starts
at a non-zero value, it will eventually approach zero 
under the influence of its potential as long as $\alpha_i$ is
an even integer.  The $\phii$ field may 
come to dominate the energy density of the Universe depending upon
the magnitude of its initial value and the speed at which it
goes toward zero, which is determined by $\lambda_i$ and $\alpha_i$.
Thus in this model, the period of vacuum dominance will not last 
forever, since eventually
$\phii$ reaches zero and then oscillates around zero. 
During the oscillation phase, $\phii$ effectively behaves as material
with ${\bar w_i} = (\alpha_i-2)/(\alpha_i+2)$, and density scaling as
$\rho_i \propto a^{-3(1+{\bar w_i})}$ \cite{turner:83},
unless it couples and decays into ordinary particles.  Thus 
any potential with $\alpha_i>4$ will eventually fade away faster
than radiation or matter unless particles are produced.

Naively, with more than one scalar field this dominance
of the ``false vacuum" can happen several times,
with first $\phi_1$ coming to dominate and then fade away, and then
$\phi_2$ coming to dominance and so on.
Since the values of $\phii(init)$, $\lambdai$, and $\alpha_i$ determine when
and how long each $\phii$ dominates, an appropriate ensemble of
such values could give a series of periods of vacuum domination
and universal acceleration, followed by periods of radiation and/or
matter domination, depending upon how each $\phii$ decays and when
the next field rises to dominance.

Very roughly the time of vacuum dominance for any $\phii$ occurs
when its energy density $\rhoi = V_i(\phii) + \frac{1}{2} \dot{\phii}^2$
equals the radiation (or matter) energy density 
$\rho_{rad}  \approx  8.6 \times 10^{-5} a^{-4} \rhocrit$.
That is, supposing for simplicity
that the scalar field kinetic energy is not large compared with
the potential energy, we have 
$a_i(vacuum~dom) \sim \left( \lambdai 
{\phii^{\alpha_i}(init)/ 8.6 \ten{-5}\rhocrit} \right)^{-1/4}$.  
Thus if the values of $V_i(init)$ are distributed
over a wide range of values so that the $a_i(vacuum~dom)$ are distributed
over epochs from the Planck
time at $a=10^{-27}$ to today at $a=1$, then our suggested scenario might
take place.  Of course there are many ways that the $\lambdai$ and
the $\phii(init)$ might be distributed to make this work, and we will
not speculate on the precise form of the distribution since we have
no understanding of it, except for possibly at two points: 
the inflaton in the very
early Universe and the current vacuum dominated epoch.

For illustration purposes we consider two simple two-field cases,
both of which satisfy the above constraints on $w$, 
fit the SN1A data as well as a cosmological constant model
with $\Omega_V = \Omega_\Lambda = 0.7$, have nearly the same distance
to CMB last scattering, and satisfy the
constraints on the fraction of vacuum energy at decoupling and BBN.
The first example, shown in 
Figure 1a, has $\alpha_1=6$, $\lambda_1=10^{-175}$, $V_1(init)=5\ten{-4}$, 
$\alpha_2=4$, $\lambda_2=10^{-125}$, $V_2(init)=10^{-8}$,
where $V_i(init)$ is given in units of $\mpcunit$ to help with numerics and
$\phii$ is found from equation (\ref{eqn:potentialeqn}).
In these units $\rhocrit = 1.6 \ten{-8}$, $\Mp = 1.94 \ten{28}$, $\phi$ has
units of ($m_{Pl}^{1/2} {\rm Mpc}^{-1/2}$), 
and $\lambda$ has units
$(m_{Pl}^{1/2} {\rm Mpc}^{-1/2})^{4-\alpha}$.
This example has
has the current accelerating expansion coming from $\phi_2$, 
but earlier, at $z \approx 1000$, $\phi_1$ started to become
important, reaching more than 30\% of the matter density at $z \sim 300$,
and then fading away.  
Today this model gives $\Omega_V=.67$ and $w=-.97$.
This example might be ruled out since the linear growth factor
calculated from matter-radiation equality is about 30\% smaller than for
the $\Omega_\Lambda=0.7$ model (about the same as for 
an $\Omega_\Lambda=0.78$ model). 
Of course the value of $V_1(init)$ could 
be reduced to lessen the size of these effects.  

The second example shows a typical period of complete vacuum dominance
in the early Universe (figure 1b).  Here the parameters used are
$\alpha_1=10$, $\lambda_1=10^{-275}$, $V_1(init)=10^{16}$, 
$\alpha_2=4$, $\lambda_2=10^{-124}$, $V_2(init)=10^{-8}$.
Here the false vacuum dominates between redshifts of $10^5$ and
$10^7$ and also near $z=0$, and is again a small contributor at 
the time of BBN and photon decoupling.
Today this model gives $\Omega_V=.65$ and $w=-.91$.
The linear growth factor is nearly the same as for a cosmological
constant model.
Note in all these plots we took the initial values of
${\dot \phi_i}(init)$ to be zero at $a_{init}=10^{-10}$, but very similar 
results obtain when
we use equipartition of kinetic and potential energy by setting
${\dot \phii}(init) = \pm(2 V_i(init))^{1/2}$.


\begin{figure}[!h]
\setlength{\unitlength}{1cm}
\centerline{\hbox{\psfig{file=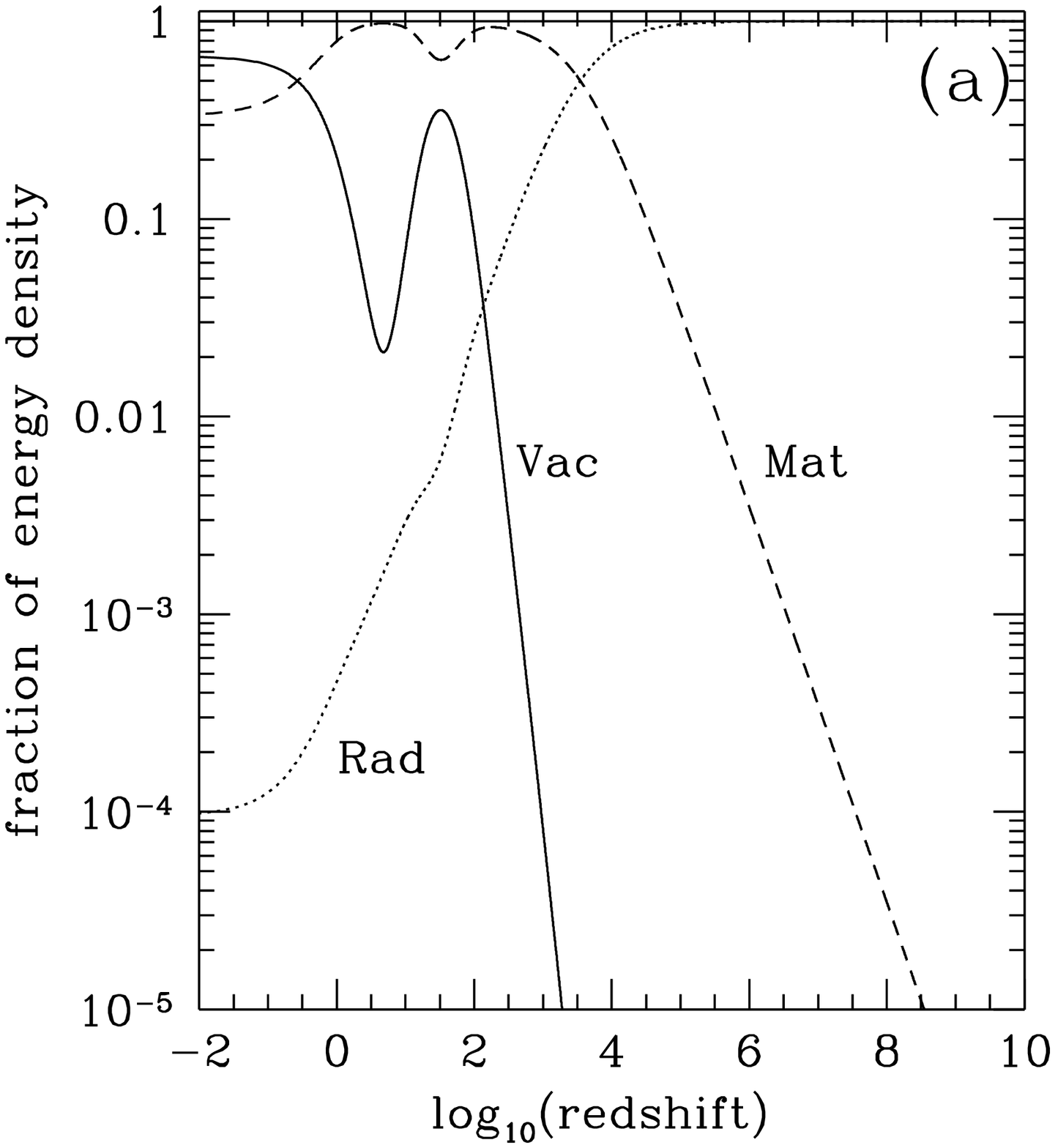,height=8cm,width=8cm}
\psfig{file=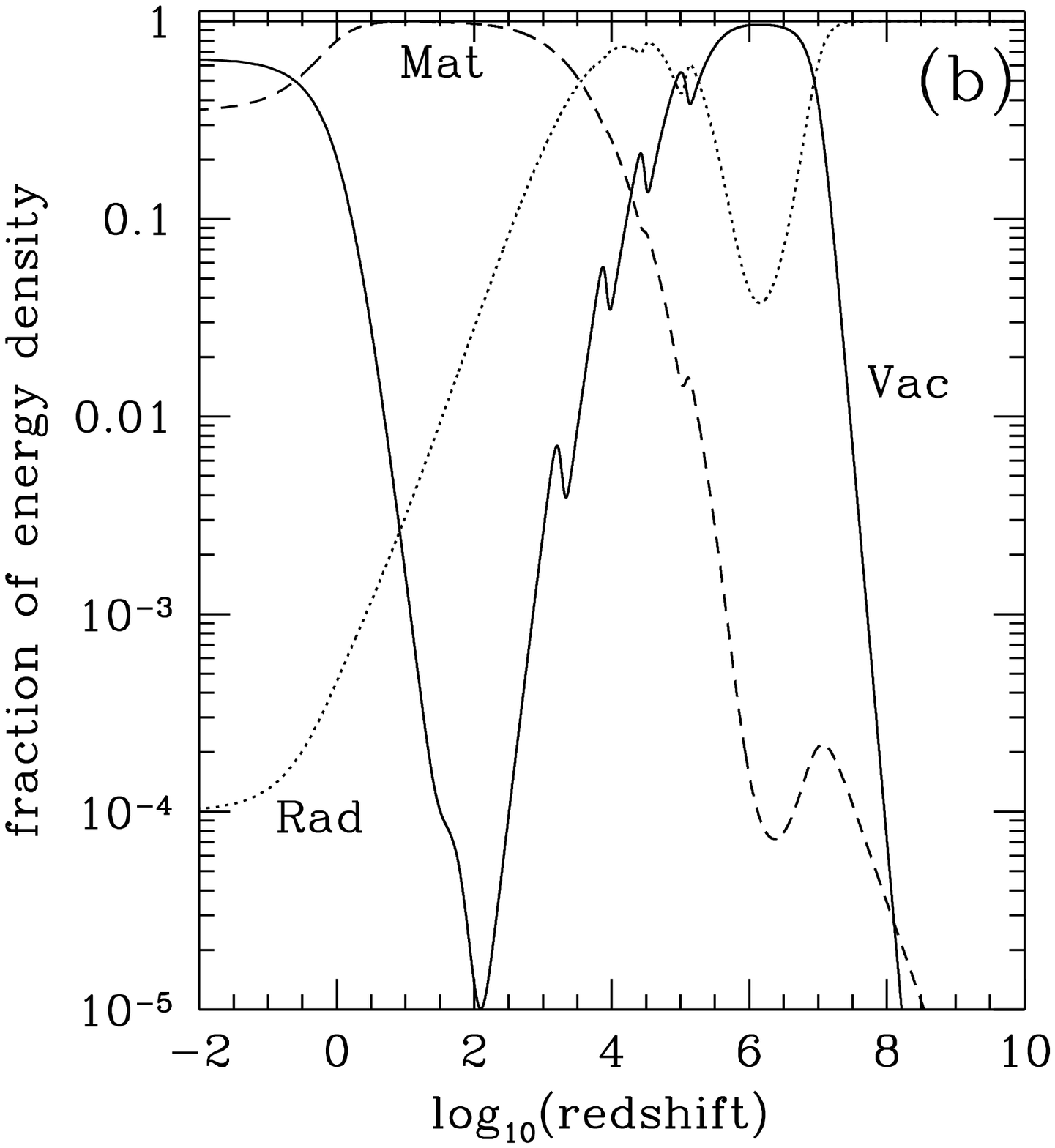,height=8cm,width=8cm}}
}
\caption{The fraction of energy density in various components vs. redshift
for simple two-field models.  
The thick dark line represents (false) vacuum energy,
the dashed line represents matter, and the thin dotted line represents
radiation.  See text for model descriptions.
}
\label{fig:examplemodels}
\end{figure}

\section*{Discussion}
This paper suggests a class of models that might explain the cosmic 
coincidence problem without invoking the anthropic principle.  The basic
idea is that there is an ensemble of scalar (or pseudoscalar)
fields with 
exponentially distributed parameters that cause them to dominate
the universal expansion at random times throughout the history of the Universe.
This paper does not attempt to solve the basic cosmological constant problem
of why the energies of all the scalar fields sum to near zero today,
but does allow for consistency between the present accelerating
expansion and a zero sum.  The simple polynomial models used here do 
require fine-tuning of their initial values and coupling constants.
The initial values determine the time
of domination and the values of the $\lambda$'s determine whether or not
vacuum domination takes place and for how long.
Note that in models with attractor potentials there is no need to fine-tune
the initial values,
and so these models have received the bulk of the attention of the community.
In our scenario,
we could probably remove this fine-tuning by considering an ensemble of
hybrid potentials which have tracking behavior at early times and then
asymptote to polynomials with $\alpha>4$ after dominance, but for 
simplicity's sake we did not pursue this option in this paper.
The basic point is that if one wants a non-anthropic solution to the
cosmic coincidence problem, one probably wants an ensemble of fields with
properties such as we discussed.  
This clue may motivate field or string
theorists to find a way of naturally generating such an unusual set of initial
conditions and coupling constants, 
or to find an ensemble of hybrid potentials that switch from
tracking to decay in the right way.

In summary, we have not investigated any models in 
detail and have no first principle
reason for why such an ensemble of fields should exist
or why their parameters should be properly distributed, but are using
the idea of a non-anthropic solution to the cosmic coincidence problem
as the main motivation.  
However, we do
note that the Higgs field in the standard model of particle physics is
a scalar field that contributes to the universal vacuum energy 
and whose contribution must be cancelled by
a large negative contribution from either a cosmological constant or from the
vacuum energy of another field (or fields).  
Including
the inflaton and the field giving rise to the
current universal acceleration we thus probably have at least three important
scalar fields contributing to standard cosmology, making our suggestion of 
a large ensemble of such fields more palatable.

This idea has several advantages, including
a ``unification" of the inflaton field and the field that currently causes
universal acceleration, as well as perhaps detectable periods of vacuum
dominance in the past, and predictions that the current era of vacuum
dominance will end and that $w$ is not precisely unity.
This last prediction might be testable in proposed experiments to
measure the value and time derivative of $w$\cite{snap}.

There are many possibilities and open questions that should be addressed.
A general question is
what kind of potentials and initial values can give rise to
realistic implementations of this idea, and what
kind of theories could give rise to such an ensemble of fields and 
initial data?   Other important questions are probably difficult to 
address without
a more specific field theoretic framework;  for example, one should
consider how each of these small periods of vacuum dominance end; 
are particles created?  
do substantial adiabatic or iso-curvature fluctuations result? 
is the Universe reheated?  Is the power spectrum affected?
Even in a model dependent way it would be interesting to
explore how much and what kind of late time vacuum domination, or near
domination, is allowed by current observations.  
It may even be that periods of vacuum dominance, or near dominance, 
could help the fit between theory and observation, or that the left over
oscillating fields could make up some of the dark matter.
In general one
could ask what sets of fields and initial data could give rise to
our current Universe and what observable effects would remain today?

\section*{Acknowledgements}
We thank Andy Albrecht, Neal Dalal, Saul Perlmutter, Ewan Stewart, Martin 
White, and Art Wolfe for discussion and valuable suggestions. 
This work was supported in part by the U.S. Department of Energy, under
grant DOE-FG03-97ER40546.

\end{document}